\documentclass[twocolumn]{pasj00}

\usepackage{natbib}

\newcommand{\fref}[1]{Figure~\ref{#1}}
\newcommand{\eref}[1]{equation~(\ref{#1})}

\def\ccm{\mbox{${\rm cm}^{-3}$}}
\def\kms{\mbox{${\rm km/s}$}}
\def\mug{\mbox{$\mu {\rm G}$}}
\def\gev{\mbox{${\rm GeV}$}}
\def\tev{\mbox{${\rm TeV}$}}
\def\yr{\mbox{${\rm yr}$}}
\def\pc{\mbox{${\rm pc}$}}

\begin{document}
\SetRunningHead{Amano~et~al.}{Stochastic Acceleration in the CMZ}
\title{Stochastic Acceleration of Cosmic Rays in the Central Molecular Zone of the Galaxy}
\author{Takanobu \textsc{Amano}, Kazufumi \textsc{Torii},
Takahiro \textsc{Hayakawa}, Yasuo \textsc{Fukui}}
\affil{Department of Physics, Nagoya University, Furo-cho, Chikusa-ku,
Nagoya, 464-8602, Japan}
\email{amanot@stelab.nagoya-u.ac.jp}

\maketitle

\begin{abstract}
Particle acceleration in the inner $\sim 200 \, \pc$ of the Galaxy is discussed, where diffuse TeV $\gamma$-rays have been detected by the High Energy Stereoscopic System (HESS) observation. The diffuse $\gamma$-ray emission has a strong correlation with molecular clouds with large velocity dispersion, indicating the presence of turbulence. It is argued that the turbulence may contribute to the acceleration of cosmic rays via stochastic acceleration. The stochastic acceleration may energize cosmic-ray protons up to $\sim 100 \, \tev$ and electrons to $\sim 1 \, \tev$ in a relatively tenuous medium. The difference in the efficiency between protons and electrons supports the hadronic scenario of the diffuse TeV $\gamma$-ray emission.
\end{abstract}


\section{INTRODUCTION}

The inner few hundreds parsec of the Galaxy or simply the Galactic center (GC) is an outstanding source of high energy radiation including $\gamma$-ray, X-ray, and non-thermal radio emission \cite[e.g.,][]{Aharonian2006,Mayer-Hasselwander1998,Koyama1996,Koyama2007,Muno2003,Law2008,Law2008a}. In addition, there exists a strong concentration of dense molecular gas in the GC. This molecular feature, the Central Molecular Zone (CMZ), consists of three major clouds, Sgr A, Sgr B2, and Sgr C \cite[e.g.,][]{Fukui1977,Scoville1975,Liszt1995}, and has a total mass of $3 \times 10^{7} M_{\odot}$ \citep{Dahmen1998}, being characterized by an unusually large velocity dispersion of $15 \, \kms$ ($1 \sigma$), and high temperature, $30 {\rm - } 300 \, {\rm K}$ \citep{Huettemeister1993,Martin2004,Oka2005}. The CMZ appears to consist of numerous dense clumps with a volume filling factor of $\sim 0.2 {\rm -} 0.5$ \citep{Sawada2001}.

\cite{Aharonian2006} reported the discovery of diffuse TeV $\gamma$-ray emission by the High Energy Stereoscopic System (HESS) from the GC. They found a correlation between the $\gamma$-ray emission and molecular clouds (MCs), which strongly indicates that the $\gamma$-rays are radiated by cosmic-ray (CR) protons via decay of neutral pions produced by nuclear collisions with ambient atoms. An important suggestion by \cite{Wommer2008} is that the spatial profile of diffuse $\gamma$-rays cannot be explained by superposition of point sources. They argued that it is likely that CRs are accelerated in the whole inter-cloud medium (ICM), otherwise the observed $\gamma$-ray emission would peak at the injection points because of relatively fast energy loss of CR protons.


\cite{Crocker2011} has recently developed a detailed steady-state model for the CR environment in the GC. They assumed that there is a fast wind ($\gtrsim 100 \, \kms$ ) blowing out from the GC, that transports most of the CR protons above and below the disk. Such a scenario seems natural in view of recently found Fermi bubbles \citep{Su2010,Dobler2010}, while so far there is no direct evidence for the wind. In any case, they found a reasonable fit to the broadband spectrum with gas densities around $n \sim 10 \, \ccm$, magnetic field strength around $B \sim 100 \, \mug$, wind speeds around $V_{\rm wind} \sim$ a few $\times \, 100 \, \kms$. The required energy input rate $\sim 10^{39} \, {\rm erg} \, {\rm s}^{-1}$ is consistent with the conventional idea that supernova remnant (SNR) shocks convert the explosion energy into CR with a conversion efficiency of $\sim 10\%$ for a supernova rate of $\sim 1/2500 \, \yr$. A possible problem in their modeling is, however, a pure power-law injection spectrum. Recent observations of SNRs in $\gamma$-ray band indicates that there is a break in the primary CR spectrum accelerated by a SNR. The break energy seems to depend on the age of SNRs, but can appear in GeV range for middle age ($\sim 10000$ yr), and in TeV range for young SNRs ($\sim 1000$ yr)  \citep[e.g.,][]{Abdo2010,Aharonian2007a} . It is argued that the spectral break may be produced as a result of the interaction between SNRs and MCs \citep{Malkov2005}. Therefore, it is natural to expect that a break will also appear in the CR spectrum injected by a supernova exploded in/around the CMZ. The $\gamma$-ray spectrum obtained by the HESS observation, however, does not show any significant spectral breaks up to $\sim 10$ TeV, indicating that the primary CR spectrum extends above $\sim 100$ TeV. If this is the case, we need a reacceleration mechanism of pre-existing CRs above the spectral break.

It is important to note that the unusually large velocity dispersion measured in the GC indicates the presence of strong turbulence. Given the fact that the peak positions of velocity dispersion coincident with the TeV $\gamma$-ray peaks \citep{Aharonian2006}, we suggest an alternative particle acceleration mechanism due to the turbulence. Namely, we discuss a possible contribution of stochastic acceleration (also known as the second order Fermi acceleration) by Alfv\'enic turbulence in the GC environment. We find that the stochastic acceleration process with parameters relevant to the GC can energize CR protons up to $\sim 100$ TeV in $\sim 10^5$ yr in a relatively tenuous medium, while the maximum energy of electrons is limited to $\sim 1$ TeV due to strong synchrotron cooling, which favors the hadronic scenario of the diffuse TeV $\gamma$-ray emission.

\section{STOCHASTIC ACCELERATION MODEL}
The original idea of stochastic acceleration proposed by \cite{Fermi1949} is to invoke random motions of magnetic clouds with typical velocities of $\sim 10$ km/s. Charged particles colliding with the clouds gain energy due to the difference between head-on and head-tail collision probabilities. It is well-known, however, that the acceleration is far too slow if we take a mean free path (or an inter-cloud distance) of $\sim 1 \, \pc$. This picture is later elaborated by considering pitch-angle and momentum-space diffusion by Alfv\'enic turbulence. The Alfv\'en wave is an incompressible, weakly damped normal mode of magnetized plasmas. Because of its nearly dissipationless feature, Alfv\'en waves are believed to be ubiquitous. The essential difference from the Fermi's original idea is that the scattering mean free path is now determined by the power of the turbulence spectrum at the resonant wavenumber, but not by inter-could distances. Note that the stochastic acceleration is invoked in many other astrophysical environments as well, including solar corona, black-hole accretion disks, lobes of giant radio galaxies, and $\gamma$-ray bursts.

In this study, we consider only spatially integrated CR spectra in which detailed geometry is assumed to be unimportant. Therefore, we are essentially considering the CR distribution in a box. The escape of CRs from the box either due to diffusion or advection is thus treated by an energy-dependent escape term with a time scale of $\tau_{\rm esc}$, i.e., it is independent of energy for the advection-dominated escape, but may depend on energy for the diffusion-dominated escape. The CR diffusion-convection equation then reduced to \cite[c.f.,][]{Petrosian2004}
\begin{equation}\textstyle
 \frac{\partial N}{\partial t} = \frac{\partial}{\partial E}
\left[
\frac{E^2}{\tau_{\rm acc}}\frac{\partial N}{\partial E} -
\left(\frac{2}{\tau_{\rm acc}} - \frac{1}{\tau_{\rm loss}}\right) E N
\right] -
\frac{N}{\tau_{\rm esc}} + Q(E),
\label{eq:diff-conv}
\end{equation}
where $E$ is the kinetic energy, $N(E,t) = 4\pi p^2 \int f(t,\mathbf{x},p) d\mathbf{x}$ with $f(t,\mathbf{x},p)$ being the distribution function, $Q(E)$ is the total injection flux, $\tau_{\rm acc}$ and $\tau_{\rm loss}$ represent typical acceleration and energy loss time scales, respectively. Note that in the above equation we use the approximation that the particle energy is highly relativistic. The important parameter for the stochastic particle acceleration is the spatial diffusion coefficient $\kappa$. Assuming elastic scattering of CRs in the frame moving with Alfv\'en waves, we may estimate the acceleration time scale as
$\tau_{\rm acc} = \left(\frac{1}{E}\frac{dE}{dt} \right)^{-1} \simeq \frac{\kappa}{v_A^2}$,
where $v_A$ represents the Alfv\'en speed \cite[e.g.,][]{Skilling1975}. Although it is possible to calculate the diffusion coefficient $\kappa$ directly by using quasi-linear theory, this theory is based on a number of assumptions which may not necessarily be valid in general. In the present study, we instead use the Bohm scaling $\kappa = \frac{1}{3} \xi r_g c \propto E$, where $c$ is the speed of light, and $r_g$ is the particle gyroradius. This expression indicates that the mean free path of the particle can be written by $\xi r_g$, meaning that $\xi$ represents the efficiency of scattering and is sometimes called the ``gyrofactor''. Numerical values of $\xi$ estimated from the solar modulation of low-energy ($\lesssim$ GeV) CRs in the heliosphere are of order $10{\rm -}100$, but may become as small as $\sim 1$ (the Bohm limit) in the vicinity of strong shocks. Given the large velocity dispersion measured in the GC, we anticipate that there exist regions in the GC where Alfv\'enic turbulence is so strong that $\xi$ becomes the same order as in the solar wind. The typical acceleration time scale can then be written using parameters relevant to the GC as
\begin{equation}\textstyle
 \tau_{\rm acc} \simeq 2.1 \times 10 \, \yr
\bigg(\frac{\xi}{10}\bigg)
\bigg(\frac{B}{100 \, \mug}\bigg)^{-3}
\bigg(\frac{n}{100 \, \ccm}\bigg)
\bigg(\frac{E}{\gev}\bigg). \label{eq:tauacc_eval}
\end{equation}
The maximum attainable energy through this process may be estimated by equating this to characteristic loss or escape time scales. We are mainly interested in CRs with energies much higher than 1 GeV. For protons of such high energies, the dominant loss process is due to nuclear collisions with ambient atoms with an energy-independent characteristic time scale of $\tau_{\rm loss, pp} \simeq 3.7 \times 10^5 \, \yr \, (n/100 \, \ccm)^{-1}$. In the case of electrons, the ionization, bremsstrahlung, inverse Compton, and synchrotron losses are potentially important. In the GC environment, since the magnetic field strength is estimated to be $\gtrsim 50 \, \mug$, and most likely $\gtrsim 100 \, \mug$ \citep{Crocker2010,Crocker2011a}, the dominant loss process is the synchrotron cooling with $\tau_{\rm loss,sync} \simeq 1.3 \times 10^6 \, \yr \, (E/\gev)^{-1} (B/100 \,\mug)^{-2}$ unless the ambient density is too high $n \gtrsim 10^{3} \, \ccm$ in which bremsstrahlung with $\tau_{\rm loss,brems} \simeq 2.4 \times 10^{5} \, \yr \, (n/100 \, \ccm)^{-1}$ becomes important. In addition, if the suggested fast wind ($\gtrsim 100$ km/s) exists \citep{Crocker2011,Crocker2011a,Crocker2011b}, a corresponding escape time scale of $\tau_{\rm esc} \simeq 4.1 \times 10^{5} \, \yr \, (V_{\rm wind}/100 \, \kms)^{-1} (h / 42 \,\pc)$ should also be taken into account, where $h$ is the height of the $\gamma$-ray emitting region. Alternatively, large scale flows over $\sim 100 \, \kms$ found in this region \citep{Fukui2006,Molinari2011} may also contribute to the escape. Since the nature of the flows does not change our argument, we shall consider only a global wind blowing out from the disk in the following discussion.

\begin{figure}[h]
\begin{center}
 \FigureFile(86mm,86mm){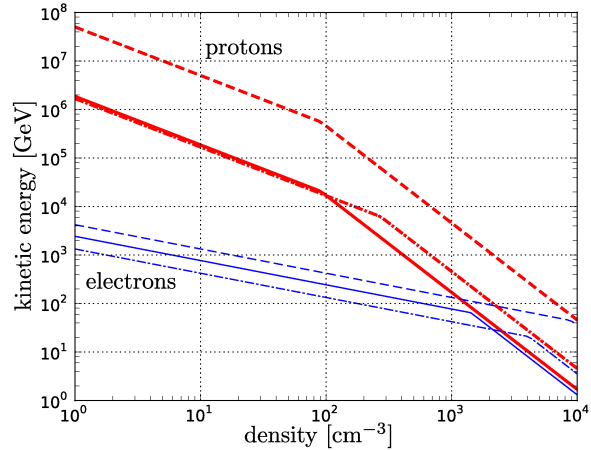}
\end{center}
 \caption{Maximum attainable energy by stochastic acceleration as a function of ambient density. Thick (upper three) and thin (lower three) lines represent the maximum energy for protons and electrons, respectively. Solid lines are for $\xi = 10$, $B = 100 \, \mug$, $V_{\rm wind} = 100 \, \kms$, while dashed lines are for $B = 300 \, \mug$ with keeping other parameters constant. Dash dotted lines are for $\xi = 100$, $B = 300 \, \mug$, $V_{\rm wind} = 300 \, \kms$.}
 \label{fig1}
\end{figure}

Combining these time scales, the estimated maximum attainable energy by the stochastic acceleration as a function of ambient density is shown in \fref{fig1}. In most cases, the maximum energy for protons is much larger than electrons because of strong synchrotron cooling which dominates at $n \lesssim 10^{3} \, \ccm$. For protons, the dominant loss changes from hadronic collisions at high density regions $n \gtrsim 100 \, \ccm$ to the convective escape by the hypothetical global outflow in lower-density regions. Even in the presence of a fast wind of a few hundreds $\kms$ that sweeps CRs out from the system in $\sim 10^{5} \, \yr$, it is possible to accelerate protons to $\sim 100$ TeV or even more when $n \lesssim 10 \, \ccm$, that is lower than the volumetric average density $n \sim 100 \, \ccm$ in that region. It is important to mention that the GC environment is highly inhomogeneous, where MCs with much higher density ($\gtrsim 10^{3} \, \ccm$) and tenuous ICM coexist. The modeling result of \cite{Crocker2011} actually favors the target gas density of $n \sim 10 \, \ccm$, suggesting that CRs accelerated in low-density regions do not penetrate deep into MCs. Therefore, if $\gamma$-rays are originating from such regions, the stochastic acceleration can provide a non-negligible contribution to the $\gamma$-ray flux.

We now investigate the allowed parameter range of the gyrofactor $\xi$ in which stochastic acceleration alone is capable of accelerating CR protons up to $\sim 100$ TeV; the constraint imposed by the HESS observation. This may have an important implication because it is generally difficult to calculate the diffusion coefficient both theoretically and experimentally in a realistic, strongly turbulent environment. For this purpose, we may rewrite \eref{eq:tauacc_eval} by equating $\tau_{\rm acc}$ with an unspecified loss or escape time scale $\tau$,
\begin{equation}\textstyle
 \xi \simeq 6.3
\bigg(\frac{B}{100 \, \mug}\bigg)^{3}
\bigg(\frac{n}{100 \, \ccm}\bigg)^{-1}
\bigg(\frac{\tau}{10^{6} \, \yr}\bigg)
\bigg(\frac{E_{\rm max}}{100 \, {\rm TeV}}\bigg)^{-1}.
\end{equation}
Since nuclear collisions are not the dominant loss in low-density regions $n \lesssim 10 \, \ccm$, $\tau$ is essentially determined by the wind velocity \citep{Crocker2011}. A wind velocity of a few hundreds $\kms$ indicates that $\xi \sim 10 {\rm -} 100$ is the necessary condition for $B = 100 {\rm - } 300 \, \mug$. These values are similar to those estimated in the heliosphere. We thus argue that there exist regions in the GC where such strong diffusion and corresponding stochastic acceleration are actually taking place. A possible example of such scenario is HESS J1745-303, which has been discussed recently by \cite{Hayakawa2011}.

The fact that the GC is far from uniform, and the CR production and radiation may occur in different environment indicates the importance of understanding the interactions between these regions. In general, the acceleration region may have a finite bulk velocity relative to MCs. It is therefore natural to consider collisions between the two distinct regions. Actually, we expect stronger hadronic $\gamma$-ray emission from such an interaction region than more tenuous (i.e., not interacting with MCs), but potentially more efficient acceleration regions. The convective transport time scale of CRs deep into a MC is estimated as $\sim 6.5 \times 10^{5} \, \yr \, (l / 10 \, \pc) (\sigma/15 \, \kms)^{-1}$, where $l$ is the typical size of a MC, and $\sigma$ is the velocity dispersion. If CRs were merely convected by the background flow, they would lose energies faster than this time scale because the hadronic-collision loss becomes increasingly significant during the convective transport. In reality, CRs can diffuse against the flow and thus remain in the acceleration region longer than expected above. In the following, we consider only the case of a one-dimensional interaction for simplicity. The diffusion length of CRs against a bulk flow $u$ can be written as $\kappa/u$. We may thus estimate the time scale of CR escape to a dense MC as $\tau_{\rm esc} \simeq \kappa/u^2$, where the acceleration is essentially terminated. It is important to notice that $\tau_{\rm acc}$ and $\tau_{\rm esc}$ have the same energy dependence, i.e., both are proportional to $\propto E$ under the Bohm scaling assumption. In this case, the resulting steady-state spectrum above the injection energy ($Q(E) = 0$) with negligible losses ($\tau_{\rm loss} = \infty$) becomes a power law. More specifically, by substituting these into \eref{eq:diff-conv} and assuming a power-law function $N(E) \propto E^{-p}$, it is easy to show that the power-law index $p$ should satisfy $p = -1 + \sqrt{1 + \frac{\tau_{\rm acc}}{\tau_{\rm esc}}}.$
The ratio of two timescales $\tau_{\rm acc}/\tau_{\rm esc} \simeq (u/v_A)^2$ thus determines the power-law index. A measured spectral index of $\sim 2.3$ requires a fast colliding velocity of $u \sim 3.1 \, v_A$ which is similar to the inferred wind velocity of a few hundreds $\kms$. It is possible to interpret that the measured $\gamma$-rays are originating mainly from such strongly-interacting regions, although we think that the theory at the present stage may oversimplify the situation for quantitative comparisons. This is partly because we assume that the acceleration region is a homogeneous medium, while interactions with MCs inevitably produce inhomogeneity, particularly gradients in density and magnetic field. Other reasons include steady-state, and one-dimensional assumptions, etc. Therefore, the present model must be taken as a qualitative model rather than a quantitative one. Nevertheless, we consider that the stochastic acceleration can in principle play a role for accelerating $\gtrsim$ TeV energy CR protons and electrons in the GC, particularly in the low-density ICM.

\section{SUMMARY AND DISCUSSION}
\cite{Aharonian2006} have proposed a time-dependent scenario in which the diffuse CRs in the GC are produced by a putative past event occurred $\sim 10^{4} \, \yr$ ago, and then diffuse over a $\sim 100 \, \pc$ scale. Recent picture of the GC is, however, such that particle acceleration to multi-TeV energies occurs not only in individual (i.e., point-like) sources but also in the whole ICM. This is because simple spatial diffusion from a single source will produce a peak at the source position, which is not consistent with the observed spatial profile. Also, no evidence for diffusion hardening due to an energy-dependent diffusion coefficient at the propagation front is reported so far.

An important indication from observations is that large velocity dispersion comparable to Alfv\'en velocity exists throughout the GC, indicating the presence of turbulent flows. The total turbulent kinetic energy may be estimated as $\sim 10^{53}$ erg, while that of diffuse CR component is $\sim 10^{49}$ erg in $4{\rm -}40$ TeV and $\sim 10^{50}$ erg if the spectrum extends from $10^{9}$ to $10^{15}$ eV \citep{Aharonian2006}. Therefore, the observed turbulence has sufficient energy to produce the diffuse CRs whatever the mechanism is. In the present study, we argue that the stochastic acceleration by Alfv\'enic turbulence can accelerate CRs up to $\sim 100$ TeV for protons, and $\sim 1$ TeV for electrons within a reasonable parameter range. The difference in the maximum energy between two components favors the hadronic scenario of the TeV $\gamma$-ray emission, consistent with the correlation found between MCs and $\gamma$-rays. Of particular note is that the GC is highly inhomogeneous; dense, but clumpy MCs are embedded in much more tenuous, volume-filling regions where the stochastic acceleration is more efficient.

\cite{Crocker2011} has also suggested that the $\gamma$-ray emitting region is less dense $\sim 10 \, \ccm$ than the volumetric average density of $\sim 100 \, \ccm$, which is basically consistent with our idea. Our theory is compatible with their assumption that the wind blowing from the GC transports most of the CR protons that illuminate the Fermi bubbles. However, the lower limit of the wind velocity determined to explain radio synchrotron emission from the GC lobe does not apply in our case. This is because the stochastic acceleration can overcome synchrotron cooling in the energy range of radio-emitting electrons. If an assumed empirical relation between the total infrared and radio-continuum (and also $\gamma$-ray) luminosities applies to the GC, a fair lower limit would be $\sim 100 \, \kms$ to sweep out CR protons before they are convected into the cores of MCs \citep{Crocker2011a}. In any case, it is likely that CRs accelerated in the ICM emit hadronic $\gamma$-rays by interacting with MCs. With this in mind, we investigate collisional interactions between the acceleration regions and MCs that may produce a power-law type spectrum. More detailed studies on this subject are still needed.


We also comment on a similar work done by \cite{Melia2011}. While their idea itself is essentially the same as the stochastic acceleration model, their equation for the energy gain is too optimistic as also noted by themselves. Their model at least differs from the standard stochastic acceleration model, which must be considered with great care.

Finally, we would like to mention that the stochastic acceleration is of course not the only candidate mechanism of CR acceleration in the GC. Magnetic reconnection and shocks (not associated with SNRs) may also contribute acceleration of CRs in the turbulent GC \citep[e.g.,][]{LeschH.1992}, which should be elaborated in the future.

\bigskip
This work is in part supported by the Global COE Program of Nagoya University (QFPU) and KAKENHI 22740118 (T.A.) from JSPS and MEXT of Japan.

\end{document}